\documentclass[12pt]{iopart}
\usepackage{graphicx}
\usepackage{epsfig}
\newcommand{\bsigma}{{\mbox{\boldmath $\sigma$}}}

\newcommand{\btau}{\mbox{\boldmath $\tau$}}
\newcommand{\bxi}{{\mbox{\boldmath $\xi$}}}
\newcommand{\btheta}{{\mbox{\boldmath $\theta$}}}
\newcommand{\bJ}{\ensuremath{\mathbf{J}}}
\begin{document}
\title{Survey propagation at finite temperature: application to a Sourlas code
as a toy model}
\author{B. Wemmenhove and H. J. Kappen}
\address{SNN, Radboud University Nijmegen \\ Geert Grooteplein 21, 6525 EZ,
Nijmegen, The Netherlands {\rm b.wemmenhove@science.ru.nl}}
\begin{abstract}
In this paper we investigate a finite temperature
generalization of survey propagation, by applying it to the problem of 
finite temperature decoding of a biased
finite connectivity Sourlas code for temperatures lower than the
Nishimori temperature. We observe that the result is a shift
of the location of the dynamical critical channel noise to larger values
than the corresponding dynamical transition for belief propagation, as 
suggested recently by Migliorini and Saad for LDPC codes. 
We show how the finite temperature 1-RSB SP gives
accurate results in the regime where competing
approaches fail to converge or fail to recover the retrieval state.
\end{abstract}
\pacs{89.70+c, 89.90+n, 05.50+q}
\section{Introduction}
In the recent past, there have been a number of developments in 
 the field of information processing and computer science that were
triggered by insights originating from statistical mechanics.
A number of them was initiated after the discovery of the close relationship
between the 
sum-product algorithm or loopy belief propagation \cite{Pearl} in
computer science on the one hand, and
the Bethe approximation, the cavity method and the replica method 
in statistical mechanics on the other hand.
\cite{YedFreeWeis,KabashSaad}. 
The belief propagation algorithm is a method to compute approximations of
the microstate
probability distribution or the corresponding marginals of variables that 
interact on a given graph. This objective arises in various statistical
inference problems with which the machine learning community is 
concerned, such as 
stereo vision \cite{SunLi}, optimization problems \cite{MeZechPa},
image restoration \cite{TanaMori}, digital zoom and shape-from-shading
\cite{FreePasCar}, error correcting codes for telecommunication 
\cite{Gallager,McElice}, human movement modelling \cite{Sigal}, 
protein folding \cite{Yanover} evaluating postitions in the game of 
Go \cite{Stern}. 
On the other hand, in the field of statistical mechanics, the main interest
is the
macroscopic behaviour of systems, and thus the objective is the calculation
of macroscopic observables describing ensemble 
averages over graphs distributed according to a given measure. In the 
limit of infinite system size, the macroscopic observables for a given 
instance of a graph coincides with the ensemble average in many models 
of interest.

The cavity method for sparse random graphs may be viewed as the macroscopic
counterpart of belief propagation. If belief propagation converges, the 
cavity method describes its macroscopic statistics in the limit of infinite
system size \cite{KabashSaad,Kabash03}.
A more careful interpretation is needed when belief propagation ceases to 
converge, and in this case more exotic statistical mechanics techniques
can play a crucial role in generalizations of the belief propagation 
algorithm. 

When belief propagation (BP) breaks down, one may distinguish between 
different possible causes for this failure. One of these possible causes is 
replica symmetry breaking, related to the breaking of ergodicity. 
This is reflected by a 
free energy landscape exhibiting a structure of a large number of
free energy valleys that are disconnected, in the sense that the correlations
between them are smaller than correlations within one valley, and
the free energy barriers between valleys are extensive.
A generalisation of the cavity method on sparse graphs has been developed to 
take into account 
a first step of replica symmetry breaking \cite{MePa01,MePa03} 
over the last years. 
Although this scheme was designed for the computation of macroscopic 
observables, i.e. averages over ensembles, it could be adapted to yield
an algorithm that gives valuable new information on specific instances of
hard optimization problems: survey propagation \cite{MeZechPa,MeZech2}.

Survey propagation (SP) has proved to be a succesful algorithm which 
can cope with
hard constraint satisfaction problems such as random 3-satisfiability 
and graph colouring \cite{BraunZech}. 
These problems become hard when
the solution space breaks apart in disconnected components, such that 
the relative sizes of the basins of attraction corresponding to these
solutions for algorithms like ordinary 
belief propagation vanish
and many suboptimal fixed points appear.
Loosely speaking, the basic idea of survey propagation is to gather the 
statistics of the
fixed points that correspond to solutions of the constraint satisfaction 
problem
\footnote{Strictly speaking, there are a number of subtle differences, such
as the assumption that the cavity fields take integer values.
Note that nonconvergence of BP is not the only problem which is cured by
SP: algorithms that are guaranteed to converge to local minima of the Bethe 
free energy in contrast to SP do not find solutions to 3-sat problems in the 
hard-sat phase
\cite{Pretti}, i.e. the local minima one arrives at are not likely to be a 
groundstate}. 
These ``surveys'' (probability distributions associated to each link 
of the graph) 
may be used for decimation of the original
constraint satisfaction problem, until the problem is reduced to a sufficiently
easy one that can be solved with a conventional algorithm.
One of the reasons why SP is so succesful is the fact that it corresponds
to a zero-temperature problem in statistical mechanics, having the consequence
that the ``surveys'', which represent probability distributions, have a very
efficient parametrization. This makes it possible to study very large graph 
instances.

The 1 step replica symmetry broken (1RSB) version of the cavity method is 
applicable to finite temperature cases as well \cite{MePa01}, suggesting
a possible potential for the algorithmic counterpart on a specific instance.
The benefit aimed at is that in cases where the free energy landscape is
complex and belief propagation itself does not converge, the surveys
provide sufficient information leading to improved marginal
approximations. 

However, various difficulties, some practical and some technical, 
must be taken into account, which we will
shortly list here, to be discussed more thoroughly later in the text:
\begin{itemize}
\item
The surveys we will be dealing with are not efficiently parametrized, since
they can take values on a continuous interval. We will be forced to
sample from distributions, which severely limits the system sizes. 
For infinite values of the connectivity of the corresponding graphs, 
more efficient representations become 
possible, as suggested in \cite{Kabashimacdma,Saadcdma}.
\item
Given that we have a limited system size, we must remember that the 
approach is likely to be sensitive to finite size effects. 
One of the issues is the optimization of the
free energy with respect to the replica symmetry breaking parameter $m$.
This may be interpreted as zooming in at the states with the lowest 
free energy, where, moreover, the ``complexity'' (entropy of states)
vanishes in a truly 1RSB scenario. It is questionable what remains of this
idea for a small system size. 
\item
It is not straightforward to test the performance of the algorithm: if
the output of the algorithm is a set of approximate 
single-spin marginal probabilities,
how can we judge the quality of this approximation? Comparing to exact results
is possible for very small system sizes only ($N=O(10^2)$). 
\item
We should take into account 
the possibility of further steps of replica symmetry breaking, which 
might prevent the algorithm from converging, 
or cause malfunction in a different manner. 
\end{itemize}

The above complications show that our expectations regarding an algorithm
based on the 1RSB generalisation of the cavity method should be modest.
However, in the current paper we present encourageing results, indicating 
that there is at least a possible
type of application, albeit for future generations of fast computers, in
the field of error correcting codes. 
This point has very recently
been recognized and suggested by Migliorini and Saad \cite{Saadrsb} for 
LDPC codes, where the authors argued that a 1RSB type algorithm 
might increase the dynamical limit of the
channel noise for decoding.
In this paper we for the first time propose
such an algorithm for the biased finite connectivity Sourlas code with
$3$-spin interactions. We show how a finite temperature survey propagation
algorithm gives
accurate results in the regime where competing
approaches fail to converge or fail to recover the encoded message.

The suggested algorithm may be used for finite temperature decoding of
encrypted messages,
where the single spin marginals are the basis for inference of original
message bit values. 
A corresponding performance measure of the algorithm is the bit error rate, 
which is easily calculated. Thereby the required knowledge of exact values of 
marginals is circumvented and one is not limited to small system sizes.
The model we will study is a biased finite connectivity Sourlas code with
$3$-spin interactions. This model is closely related to a general $3$-spin
model with external fields.

In the next sections we will introduce the general model class of Ising
$3$-spin systems with external fields, we will shortly
review belief propagation, the cavity method and it's 1RSB generalization,
and finally describe the finite temperature generalization
of survey propagation.
Exact comparison
of marginals is only possible for limited system sizes ($N=45$ using a
junction tree algorithm \cite{Pearl}), 
which indeed turned out to be too small to 
see any improvement in a few cases we studied. Therefore 
we consider, as a toy model, 
finite temperature decoding of
the finite connectivity Sourlas error correcting code based on 3-spin 
interactions. As was also pointed out for the case of 
a CDMA model in \cite{Kabashimacdma,Saadcdma}, 
when decoding happens without the knowledge of the channel noise, one may
suffer from replica symmetry breaking effects if the adopted estimate of the 
channel noise is below the Nishimori temperature \cite{Nishimori}.
The appearance of metastable states disrupts the behaviour
of simple message passing algorithms before the onset of replica symmetry
breaking of the equilibrium state \cite{FranzLeoneetal}. 
As mentioned above, in \cite{Saadrsb} it was
proposed that a 1RSB type algorithm might be able to increase the critical
noise value for the decoding dynamics as compared to RS type decoding 
algorithms.

\section{The $3$-spin model}
Although the formalism may be set up in a more general fashion, we will be 
focussing on systems of $N$ binary variables $\sigma_i \in \{-1, 1\}$, 
$i \in \{1,2,\ldots,N\}$ that interact on a random graph, where the order
of the interaction is $3$. The equilibrium microstate probability distribution
will be described by the Boltzmann measure 
\begin{equation}
p(\bsigma) = \frac{\exp[-\beta H(\bsigma)]}{\sum_{\btau} \exp[-\beta H(\btau)]}
\end{equation}
where the Hamiltonian is given by
\begin{equation}
H(\bsigma) = \sum_{\nu} J_{\nu} \sigma_{\nu_1} \sigma_{\nu_2} \sigma_{\nu_3} 
+ \sum_i \theta_i
\sigma_i
\label{ham}
\end{equation}
where $J_{\nu} \in \{-1, 1\}$.
Greek letters $\mu,\nu$ will label interaction vertices (factors in a 
factor graph representation), $\nu_i$ label spins in $V(\nu)$, the set of
site indices contained in interaction $\nu$.
We will be discussing random graphs with fixed connectivity, i.e. the
number of interactions for spin $i$, $|V(i)|$,
will be fixed and equal for all $i$, where $V(i)$ is the set of all 
interactions containing $i$, i.e., we will choose $|V(i)| = k \ \forall i$ 
from now on. Consequently, the sum over $\nu$ in the Hamiltonian
(\ref{ham}) runs from $1$ to $kN/3$, the total number of interactions.

In a statistical inference problem, we assume the values of $\{ J_{\nu} \}$, 
$\{ \theta_i \}$ and $\beta$ are known, and the objective is the calculation
of the partition sum, or of a marginalized probability, e.g.
\begin{equation}
p(\sigma_i) = \sum_{\bsigma \setminus_i} p(\bsigma)
\end{equation}
In general the calculation of this quantity requires $O(2^N)$ operations, and
thus for large $N$ one resorts to approximations. 

In the regime of finite connectivity, the average loop length is of
order $\log(N)$, such that for large system sizes, correlations that 
propagate through loops may be neglected if the couplings are of order
$N^0$. In this regime the Bethe 
approximation, which is exact on tree graphs, performs well. Consequently,
the cavity method gives rather accurate results, as long as the system
is in a single state, and belief propagation is a good candidate for
inference on a (large) single instance. \cite{Kabash03}.
For large values of $\beta$ (low temperature values), replica symmetry
breaking occurs, and a single state cavity method 
description is no longer accurate. 
Belief propagation ceases to converge, or gets trapped in a sub-optimal
free energy minimum \cite{FranzLeoneetal}, and does not 
give any information. 
In the absence of external fields, it has been shown that the 
3-spin model with interactions
$J_{\mu} \in \{-1, 1\}$ is self-consistently described within the 1RSB
cavity framework \cite{Franzetalpspin}, for a fixed number of
interactions per spin.

\section{The methods}

\subsection{The cavity method}
The cavity method \cite{MPV,MePa01} in it's simple form is based on two 
assumptions:
\begin{itemize}
\item
The values of the spins $\{ \sigma_j \}$ that are neighbours on the graph
of site $i$ are only nontrivially correlated through the interaction 
$\nu \in V(j)\cap V(i)$ with 
site $i$. 
\item
In equilibrium, the system is in a single state (there is just one
dominant free energy valley), i.e. there is a one to one
correspondence between the states of neighbouring spins 
\end{itemize}
As a consequence of these assumptions, in the absence of their
interactions $\nu$ with spin $\sigma_0$, 
we may write the probabilities
for $2(k-1)$ neighbours 
$\{ \sigma_j\}$ of $\sigma_0$ as independent, parametrized in the following 
way:
\begin{equation}
p_j^\nu(\sigma_j) \sim \exp(\beta h_j^\nu \sigma_j)
\end{equation}
Then linking the $2(k -1)$ spins to spin $0$, we may write
\begin{equation}
\hspace*{-20mm}
p_0^\mu(\sigma_0) \sim \sum_{\{ \bsigma_{\setminus_{\sigma_0}}\} }
\exp\left\{\beta \sigma_0 (\sum_{\nu \in V(0)_{\setminus_\mu}} [J_{\nu} 
\sigma_{\nu_2} \sigma_{\nu_3} + \theta_0)]) + \beta \left[\sum_{\nu \in 
V(0)_{\setminus_\mu}}[ h_{\nu_1}^\nu \sigma_{\nu_1}+h_{\nu_2}^\nu 
\sigma_{\nu_2}]\right]\right\}
\end{equation}
It then follows that in this state the corresponding field $h_0$ 
associated to spin $0$ (which again misses an interaction with one other
pair of spins, interaction $\mu$) is given by
\begin{equation}
h_0^\mu = \sum_{\nu \in V(0)_{\setminus_\mu}} 
\frac{1}{\beta}\tanh^{-1}[\tanh(\beta J_{\nu})\tanh(\beta h_{\nu_1}^\nu)
\tanh(\beta h_{\nu_2}^\nu)] + \theta_0
\label{cavfielditer}
\end{equation}
On a sparse random graph with random interaction values, the macroscopic
statistics will be governed by the distribution of cavity fields $W(h)$ over
the graph, which, due to the statistical independence of cavity spins, follows
from iterating equation (\ref{cavfielditer}) in distribution:
\begin{eqnarray}
\hspace*{-15mm}
W(h) =  \int \prod_{i=1}^{k-1}[{\rm d} h_i {\rm d} g_i W(h_i) W(g_i)]
\nonumber \\ 
\times
\left\langle \delta\left\{h-\frac{1}{\beta} \sum_{i=1}^{k-1} 
\tanh^{-1}[\tanh(\beta J)\tanh(\beta h_i)
\tanh(\beta g_i)] + \theta\right\} \right\rangle_{J,\theta}
\end{eqnarray}
where the brackets denote averages over the interactions $J$ and random
external fields $\theta$.
This equation may be solved numerically with a `population dynamics' algorithm
\cite{MePa01}.
From the stationary distribution under this iteration algorithm, all
macroscopic quantities can be derived.
\subsection{Belief propagation}
Belief propagation in essence is the above procedure applied to a specific
instance of a (sparse) graph. Now the procedure is focussed on the computation
of local quantities, 
all the cavity fields associated to the graph, i.e. one for each link. 
By iterating equation (\ref{cavfielditer}) for a given graph, one avoids 
averaging over the interaction values $J$ and the external fields $\theta$.
If the algorithm converges to a fixed point, from the resulting cavity 
fields one may compute the single node ``beliefs'' 
\begin{equation}
b_i(\sigma_i) = \frac{1}{2}(1+m_i\sigma_i)
\end{equation}
which are approximations for the marginal probabilities
\begin{eqnarray}
b_i(\sigma_i) \simeq p(\sigma_i) \nonumber \\
\end{eqnarray}
Their expressions in terms of the cavity fields are
\begin{eqnarray}
m_i = \tanh(\beta H_i)
\end{eqnarray}
where
\begin{equation}
H_i = \sum_{\mu \in V(i)}\frac{1}{\beta}
\tanh^{-1}[\tanh(\beta J_{\mu})\tanh(\beta h_{\mu_1}^\mu)
\tanh(\beta h_{\mu_2}^\mu)] + \theta_i
\end{equation}
The 3-spin beliefs, approximations for the 3-spin marginals are 
computed from the cavity fields in a similar way.\\
Belief propagation has been rediscovered several times,
and has a close relation with the Bethe approximation. Fixed points of
belief propagation are minima of the Bethe free energy, which is exact on 
tree-like graphs, but is often a good approximation when a graph contains
loops \cite{Yedfreeweiss05}.


For lower temperatures and for graphs with interactions inducing frustration
effects, the cavity method and belief propagation break down. Belief 
propagation often ceases to converge in these cases.  
One would expect a similar breakdown of the cavity method in this case, 
but since the order parameter of the cavity method is a macroscopic 
object, the behaviour is quite different generally. The distribution of 
cavity fields may still be stationary under the population 
dynamics iteration in parameter regimes where belief propagation does not
converge. However, this stationary distribution obviously does not represent
the distribution of locally consistent beliefs. Thus
the interpretation of this result is a more complicated issue in terms
of local quantities\footnote{From a global point of view the equations are  
equivalent to the replica symmetric approximation of the macroscopic 
observables}.
It is suggested \cite{Kabash03} 
that stability analysis through the iteration of 
perturbations in the distribution might reveal the location of the
AT line \cite{AT}, at which the replica symmetric approximation becomes
unstable. For the SK model \cite{SK} on a dense graph this has been
shown \cite{Kabash03} to be the case indeed. 
Several heuristic \cite{Jorisestimates} or rigorous statements
\cite{Ihler,Jorisuai} have been made concerning 
convergence of belief propagation in
relation to the location in the phase diagram of Ising spin type systems.

The cavity method has been extended to deal with cases in which replica
symmetry breaking occurs, and thus it is worthwile to modify this method and
apply it to the case of single instance statistical inference. In the
same manner, survey propagation and its generalization for nonzero 
energy density \cite{ZechBattspy} were developed for optimization problems. The
main difference in the following discussion will be the application to
problems at nonzero noise-level, i.e. at finite temperature.

We will note here explicitly that this approach will not cure problems
arising from short range correlations e.g. in regular lattices. 
A possible approach for this type of problems, 
which we shall not discuss here, 
is to take into account correlations between larger numbers
of spins in the free energy. This idea was first introduced by Kikuchi 
\cite{Kikuchi} 
and leads to the cluster variation method \cite{Pelizolla}

\subsection{One step replica symmetry breaking}

The cavity method at the level of one step replica symmetry breaking is
usually applied at zero temperature 
\cite{MePa03,MeZech2,RicciCastellani,MontanariParisiRicci}. 
The reason for this is that
at $T=0$ on the one hand the equations simplify, and on the other hand there
are more known applications in optimization problems for which a single
instance generalization of this 
technique is fruitful (e.g. 3-sat, graph colouring).

Here we will be applying the finite temperature version of the
cavity method, of which we will
just discuss the basic assumptions, and refer the reader to
\cite{MePa01} and \cite{MePa03} for a more thorough description.
The main assumptions, to be compared with the assumptions of the cavity
method in it's simple (replica symmetric) form, are summarized in the 
following:

In contrast with the `single state' cavity method, one now assumes that
an exponential number of states contribute to the Boltzmann measure. 
At each link of the graph, instead of regarding just one cavity field, one
takes into account a distribution of cavity fields.
The density of states at a given free energy is of an exponential form:
\begin{equation}
\rho(F) = \exp[m \beta (F-F_{\rm R})]
\label{distfreeenergy}
\end{equation}
where $F_{\rm R}$ is the reference energy and $m$ is the so-called
replica symmetry breaking parameter. This assumption is shown to be 
consistent with the iteration of cavity field distributions, given that the 
value of the cavity field in a given state is uncorrelated with the free 
energy of that particular state. A consequence is that at each iteration the
distribution is reweighted by a factor $\exp[-m\beta \Delta F]$, where
$\Delta F$ is the difference between the free energies of the state before
and after the iteration. 
Observables will first be summed over all states $\alpha$ weighted by their 
corresponding Boltzmann factors 
$\exp(-\beta F^\alpha)/\sum_\gamma \exp(-\beta F^\gamma)$, the result
of which should be averaged over the distribution of free energies 
(\ref{distfreeenergy}).

In practice, the calculation of observables is numerically implemented
by an advanced population dynamics algorithm, in which $N$ populations
of $M$ fields, each representing a distribution at a given link, are
iterated. This procedure can be executed for all values of
$m \in [0,1]$, yielding a generalized free energy $F(m)$. 
Selecting the equilibrium state is achieved by demanding
$\partial F(m)/\partial m =0$ \cite{MPV}, 
which is equivalent to zooming in at the
free energy of states for which the entropy of states (or ``complexity'') 
becomes non-extensive \cite{Monasson}. 

The numerical cost of the procedure described above is considerable, 
since the cavity field distributions are represented in terms
of (large) sample populations. 
Moreover, the extremization of the free energy with
respect to $m$ may difficult, when the variation of $F(m)$ as a 
function of $m$ is small. 
At zero temperature, for a large model class, the situation simplifies a 
lot, since the representation of cavity field distributions is much more
efficient. In the case of discrete values for the local Hamiltonian
(e.g. any spin glass with $\pm J$ interactions, k-sat energy terms) the
cavity fields can be assumed to take integer values, such that the
only degrees of freedom are coefficients of a small number of delta peaks.
For random $k$-satisfiability, a single number parametrizes the 
field distribution for each link on the graph. In the SAT phase, 
where the
energy is zero, the reweighting of distributions by the free energy shift
as described above effectively either leaves the distribution invariant,
or rules out the state completely (contradictory states), leading to the
very efficient algorithm of survey propagation on single instances of
random graphs representing the satisfiability problem.

\section{Finite temperature survey propagation}

The goal of this paper in general is an investigation of the benefits of 
a survey propagation algorithm for finite temperature.
The type of problem we address is the inference of marginal 
probabilities for a spin-glass type model displaying replica symmetry
breaking effects. Along the lines of the one step RSB cavity method for finite 
temperature we iterate
distributions of cavity fields, where instead of sampling external fields
and interaction values at each iteration, we let the distributions evolve
on a given instance of a sparse graph, thus fixing the value of $\theta$
and $J$ locally. 
For a given value of the replica symmetry breaking parameter, the 
iteration equations are the following:
\begin{eqnarray}
\hspace*{-15mm}
W_0^{\mu}(h_0^\mu)  = \int \prod_{\nu \in V(0)_{\setminus_\mu}}[{\rm d} 
h_{\nu_1}^\nu W_{\nu_1}^\nu(h_{\nu_1}^\nu)
{\rm d} g_{\nu_2}^\nu W_{\nu_2}^\nu(g_{\nu_2}^\nu)]  
e^{-m \beta \Delta F(\{ J_\nu \}, \{ h_{\nu_1}^\nu \}, \{ g_{\nu_2}^\nu\})}
\nonumber \\
\hspace*{-10mm}
\times \delta \left\{ 
h_0^\mu - \sum_{\nu \in V(0)_{\setminus_\mu}} \frac{1}{\beta}
\tanh^{-1}[\tanh(\beta J_\nu)\tanh(\beta h_{\nu_1}^\nu)
\tanh(\beta g_{\nu_2}^\nu)] - \theta_0
\right\}
\end{eqnarray}
where
\begin{eqnarray}
\hspace*{-15mm} \Delta F(\{ J_{\nu}\}, \{ h_{\nu_1}^\nu\}, \{ g_{\nu_2}^\nu\})
= \log 2 + \sum_{\nu\in V(0)_{\setminus_\mu}} \log \left\{
\frac{\cosh(\beta J_\nu)}{\cosh(\beta u(J_\nu, h_{\nu_1}^\nu, g_{\nu_2}^\nu))} 
\right\}
\nonumber \\
 + \log \cosh\left\{ \beta \left[
\sum_{\nu \in V(0)_{\setminus_\mu}} u(J_\nu, h_{\nu_1}^\nu, g_{\nu_2}^\nu) 
+ \theta_0
\right]\right \}
\end{eqnarray}
and
\begin{equation}
u(J_\nu, h_{\nu_1}^\nu, g_{\nu_2}^\nu) = \frac{1}{\beta}
\tanh^{-1}[\tanh(\beta J_\nu)\tanh(\beta h_{\nu_1}^\nu)
\tanh(\beta g_{\nu_2}^\nu)]
\end{equation}
From the distributions of cavity fields, one may construct distributions
of ``belief'' fields:
\begin{eqnarray}
\hspace*{-15mm}
\hat{W}_0(H_0) = \int \prod_{\mu \in V(0)}[{\rm d} h_{\mu_1}^\mu W_{\mu_1}^\mu
(h_{\mu_1}^\mu) {\rm d} g_{\mu_2}^\mu W_{\mu_2}(g_{\mu_2}^\mu)]  
e^{-m \beta \Delta \hat{F}(\{ J_\mu\}, \{ h_{\mu_1}^\mu \}, \{ g_{\mu_2}^\mu
\})}\nonumber \\
\hspace*{-10mm} \times \delta \left\{ 
H_0 - \sum_{\mu \in V(0)} \frac{1}{\beta}
\tanh^{-1}[\tanh(\beta J_\mu)\tanh(\beta h_{\mu_1}^\mu)\tanh(\beta 
g_{\mu_2}^\mu)]
-\theta_0 \right\}
\end{eqnarray}
where
\begin{eqnarray}
\hspace*{-15mm}
\Delta \hat{F}(\{ J_\mu \}, \{ h_{\mu_1}^\mu\}, \{ g_{\mu_2}^\mu\})
= \log 2 + \sum_{\mu\in V(0)} \log \left\{
\frac{\cosh(\beta J_\mu)}{\cosh(\beta u(J_\mu, h_{\mu_1}^\mu, g_{\mu_2}^\mu
))} \right\}
\nonumber \\
 + \log \cosh\left\{ \beta \left[
\sum_{\mu \in V(0)} u(J_\mu, h_{\mu_1}^\mu, g_{\mu_2}^\mu) + \theta_0
\right] \right\}
\end{eqnarray}
The marginal probability $p(\sigma_0)$ for the value of the spin at 
site $0$ is now given by
\begin{equation}
p(\sigma_0) = \frac{1}{2}[1+\sigma_0\int {\rm d} H_0 \hat{W}_0(H_0) 
\tanh(\beta H_0)]
\end{equation}
Following \cite{MePa01}, we can deduce expressions for the free energy of 
the system $F(m)$ and the derivative of the free energy 
$\partial F(m)/\partial m$. These are given in (\ref{appfreee})

\section{The $3$-spin model as a biased finite connectivity Sourlas-code 
for a binary symmetric channel}

The 3-spin model with finite connectivity may be regarded as an
error correcting code of the Sourlas-type \cite{Sourlas}. The Sourlas 
code for p-spin interactions was originally studied for a fully connected 
model. In the limit $p\to \infty$ it is equivalent to the random energy 
model \cite{REM} and the Shannon limit is approached. 
The benefit of looking at finite connectivity codes and 
finite $p$ is the fact that the transmission rate, which vanishes for
the dense graph and $p \to \infty$, is now finite \cite{Vicenteetal}. 
However, the drawback is the
fact that the bit error rate does not vanish, such that for practical
purposes its potential is limited. Codes of the Gallagher type (low density
parity check codes) \cite{LDPC} are more promising from that perspective, 
since they have vanishing bit error rate and approach the Shannon limit.

The optimal decoding temperature for error correcting
codes can be shown to coincide with the
temperature corresponding to the ``channel noise'' \cite{Nishimori}. 
Since this temperature is equal to the so-called ``Nishimori temperature'',
it can be shown (under certain conditions, including the absence of
external fields or ``biases'') that this temperature always corresponds to a
replica symmetric phase in the phase diagram \cite{Nishisherring}, 
suggesting that belief propagation may be an appropriate decoding algorithm. 

In real life, however, the channel noise might be an unknown quantity.  
If the estimate of the channel noise in this case
is too small, the decoding is likely to happen in a parameter region where
RSB effects play a role, 
in which BP either does not converge, or gives bad results.
One would expect that in this case, the 1RSB generalisation might be 
a robust alternative \cite{Saadcdma,Kabashimacdma}

A more detailed inspection of the generic behaviour of
error correcting codes
tells us that decoding problems are caused by the so-called dynamical
phase transition \cite{FranzLeoneetal}:
in a low temperature regime
above some critical channel noise value $p_d$, the
fragmentation of the free energy landscape prevents the algorithm from 
finding the ferromagnetic state. Above an even larger noise value
$p_c>p_d$, the free energy of a replica symmetry broken spin glass state
is lower than the ferromagnetic free energy, such that decoding is not 
possible at all. Thus, the only region in which a 1RSB algorithm might 
improve the decoding performance, is in the region $p_d < p < p_c$, 
thereby shifting the dynamical
phase transition to a higher value.

In the presence of external fields, the model may be interpreted as
a \emph{biased} Sourlas-code. The location in the phase diagram where
improvement upon conventional decoding is to be expected, is close to
the critical channel noise, for low decoding temperature. Since in 
the case of a random bias we might not have a sharp phase transition, 
we will have to be careful in the selection of a 
suitable parameter regime.

In order to select a parameter regime in which our algorithm might perform
better than BP, we will perform a short macroscopic analysis of the model.
We will here shortly discuss the cavity analysis of
the model for $3$-spin interactions and random external fields. In
\cite{Vicenteetal} the model was studied for uniformly biased messages
(a uniform external field) using the replica method. 
In principle, their result applies to the present case of random biases,
due to gauge symmetry, for the appropriate
choice of a message bias, but for clarity we will here 
discuss the derivation of this result to make the connection explicit.

\subsection{RS analysis of biased code for $3$-spin interactions}
First of all, a locally biased message $\bxi \in \{-1,1\}^N$ 
is generated, according to
\begin{equation}
p(\bxi|\btheta) = \prod_i \frac{\exp[\beta_p \theta_i \xi_i]}{2\cosh[\beta_p 
\theta_i]}
\end{equation}
The external fields or biases will be generated from the distribution
\begin{equation}
p(\btheta) = \prod_i p(\theta_i)
\end{equation}
where we choose
\begin{equation}
p(\theta_i) = c(+) \delta(\theta_i - \theta) + c(-) \delta(\theta_i
+ \theta) + c(0) \delta(\theta_i)
\end{equation}
with $\theta$ constant and of 
course $c(+) + c(-) + c(0) = 1$. \\
Given the message, the code $\bJ \in \{-1,1\}^{Nk/p}$ 
is generated according to
\begin{equation}
p(\bJ|\bxi) =
\prod_\mu  \frac{\exp[\beta_p J^\mu \xi_{\mu_1}\xi_{\mu_2}
\xi_{\mu_3}]}{2\cosh(\beta_p)}
\end{equation}
Note that the above choices correspond to a binary symmetric channel (BSC),
for which each transmitted bit is likely to be flipped to its inverse 
with equal probability $p_{\rm ch}$ given by 
\begin{equation}
\beta_p = \frac{1}{2} \log{\frac{1-p_{\rm ch}}{p_{\rm ch}}}
\end{equation}
The original message is inferred according to 
\begin{equation}
\hat{\xi}_i = {\rm sgn}(\langle \sigma_i \rangle_\beta)
\end{equation}
where the average is taken with respect to the Boltzmann measure
\begin{equation}
p(\bsigma) = \frac{\exp[-\beta H(\bsigma)]}{ \sum_{\btau} 
\exp[-\beta H(\btau)]}
\end{equation}
Thus, for the finite temperature decoding at inverse temperature
$\beta$, the knowledge of the marginal $p(\sigma_i)$ is required.
Since we want to describe the 
macroscopic behaviour of the code, we use the cavity method to average
over the disorder \footnote{Note that using the replica method in the replica
symmetric framework one generally arrives at identical equations}
(the messages $\bxi$ and the interaction values $\bJ$).

Here we summarize the equations to be iterated using a 
population dynamics algorithm (see \ref{appsourlas} for details)
\begin{eqnarray}
\hspace*{-15mm}
W(h)  = \sum_\alpha q(\theta_\alpha)
\int \prod_{i=1}^{k-1}[d u_i Q(u_i)]
\delta[h-\sum_{i=1}^{k-1} u_i - \theta_\alpha] \nonumber\\
\hspace*{-15mm}
Q(u)  = 
\sum_J  \frac{e^{\beta_p J}}{2 \cosh(\beta_p)} 
\nonumber \\ 
\times ~ \int dh dg W(h) W(g)
\delta\left[u - \frac{1}{\beta}\tanh^{-1}[\tanh(\beta h)\tanh(\beta g)
\tanh(\beta J)]\right] \nonumber \\
\hspace*{-15mm}
\hat{W}(H)  = \sum_\alpha q(\theta_\alpha) \int \prod_{i=1}^k[d u_i Q(u_i)]
\delta[H-\sum_{i=1}^k u_i - \theta_\alpha] \nonumber \\
\hspace*{-15mm}
\mu  = \int dH \hat{W}(H) {\rm sgn}(H) 
\label{caviter}
\end{eqnarray}
where
\begin{equation}
q(\theta_\alpha) = c(\alpha)\frac{e^{\beta_p \theta_\alpha}}{\cosh(\beta_p 
\theta_\alpha)}
\end{equation}

From these equations we compute the bit error rate $\rho = (1-\mu)/2$ 
as a function of 
the channel noise, 
for connectivity $k=4$ at inverse temperature $\beta = 5$, 
where we initialize the population dynamics 
algorithm with random positive fields uniformly on the interval 
$0\leq h \leq 3$,
or randomly on a symmetric interval $-3 \leq h \leq 3$.

Until now, we did not discuss the need for the bias explicitly. 
It is well-known that an unbiased Sourlas code has such a small basin of 
attraction, that BP does not converge to the ``ferromagnetic'' or
``retrieval'' state for any
order of the interaction larger than $2$, unless prior knowledge is taken into 
account in the initialisation \cite{Vicenteetal}. 
Therefore, we have added random external fields, or random biases to a 
fraction of the sites, which have the effect of breaking the local symmetry.

\begin{figure}
\begin{picture}(400,170)
\put(80,0){\includegraphics[height=6cm, width=8cm]{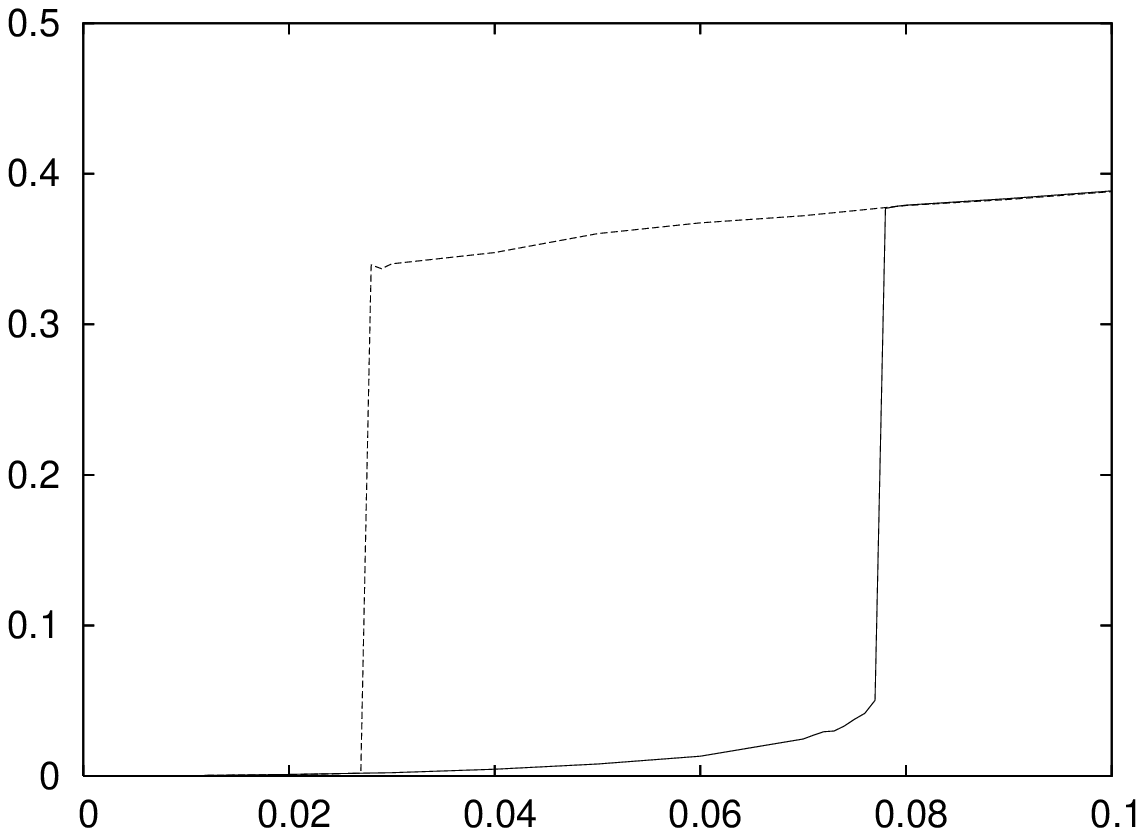}}
\put(65,90){$\rho$}
\put(205,125){SG}
\put(205,25){F}
\end{picture}
\begin{picture}(400,170)
\put(80,5){\includegraphics[height=6cm, width=8cm]{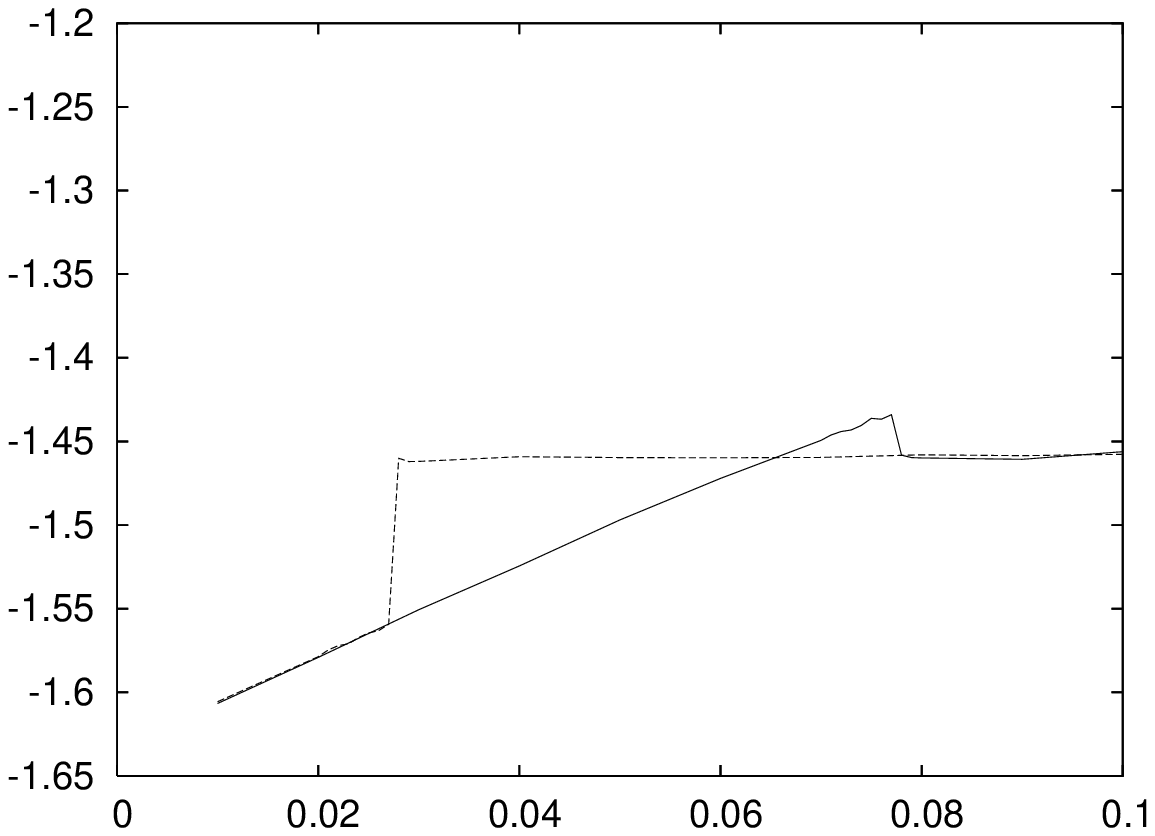}}
\put(65,95){$f_{\rm RS}$}
\put(195,0){$p_{\rm ch}$}
\put(175,90){SG}
\put(185,50){F}
\end{picture}
\caption{Top: bit error rate $\rho$ as a function of channel noise for a 
code with
$k=4$, $\beta=5$  and bias $\theta=1.5$ with ferromagnetic initial
conditions versus spin-glass initial conditions. Bottom: The RS free energies
corresponding to the top picture. In both figures the label F denotes 
ferromagnetic initial conditions and SG denotes the branch with
spin-glass type initial
conditions (see text).}
\label{biterrorversusp}
\end{figure}

We chose the fraction of external
fields according to the values $c(0) = 0.8$, $c(+) = c(-)= 0.1$, and
$\theta = 1.5$, such that the magnitude at low temperatures 
might compete with the local interaction
terms at sites where $\theta\neq 0$. 
For a temperature sufficiently below the Nishimori temperature 
($\beta_{\rm Nishimori}=\beta_p$), 
i.e. $\beta = 5$, results are reported in figure \ref{biterrorversusp}. 
Clearly there is a jump in the bit error rate as a function of the 
channel noise around $p_{\rm ch}=0.028$ for ``spin glass'' initial 
conditions, and around $p_{\rm ch}=0.78$ 
for ferromagnetic initial conditions. We
may interpret the former line as a dynamical phase transition for BP. 
Comparing the free energy values in both cases, we find that the replica
symmetric approximation of the spin-glass free energy is nearly constant
over a large interval, whereas the ferromagnetic free energy (which is 
expected to be replica symmetric indeed) is an increasing function up to the
point where the population dynamics collapse to the spin-glass state. 
The RS critical channel noise corresponds to 
the point where the two lines cross,
suggesting that retrieval is not at all possible above this point, around
$p_{\rm ch} = 0.065$. Within
the RS framework, therefore, the jump at $p_{\rm ch}=0.78$ is not the physical
critical channel noise.

\begin{figure}
\begin{picture}(400,180)
\put(80,5){\includegraphics[height=6cm, width=8cm]{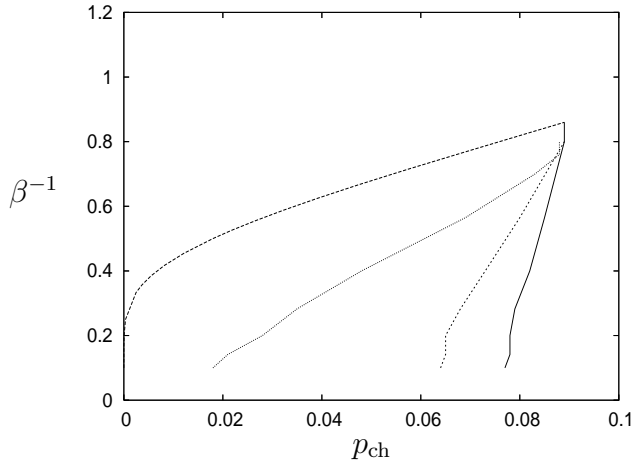}}
\put(65,95){$\beta^{-1}$}
\put(195,0){$p_{\rm ch}$}
\end{picture}
\caption{Replica symmetric phase diagram of the finite connectivity Sourlas 
code for $k=4$, $p=3$, $c(0)=0.8$, $c(+)=c(-)=0.1$, $\theta=1.5$. From top
to bottom the lines indicate the Nishimori temperature, the dynamic transition,
the RS critical transition and the line along which the population dynamics 
algorithm ceases to find the ferromagnetic state.}
\label{phasediag}
\end{figure}

Figure \ref{phasediag} displays the location of the various transition points
for a range of temperatures below the Nishimori temperature within the 
here described RS framework.
Within a 1RSB framework the results are likely to change, since in general the
spin-glass free energy will be higher than the RS approximated result.
From the above RS analysis, we expect BP to fail for $p_{\rm ch}$ above the
dynamical transition line, and we expect improvement using the 1RSB treatment
above the RS dynamical transition.





\section{Numerical results for single instance decoding}
We have tested a number of algorithms on single instances of a Sourlas code 
corresponding to the same parameter values as above, i.e. with $\beta=5$, 
way below the Nishimori line in the critical region.
Ordinary BP was able to find the ferromagnetic state roughly up to the value
of $p=0.028$, the point where the randomly initiated population dynamics
algorithm ceases to converge. Beyond that point, BP ceased to converge too,
even when we took into account the prior knowledge corresponding to the
external field in the initialisation.
However, it turned out that a ``damped'' version of BP (i.e. at each
timestep $t+1$ replacing the field $h_t$ by 
$(1-\epsilon)h_t + \epsilon h_{t+1}$) improved the performance dramatically.
In case this algorithm did not converge either, we applied a 
time averageing procedure, along the lines of \cite{Jort,Jortsaad}.
The resulting algorithm was able to find an optimal ferromagnetic-type 
state (i.e. with bit error rate comparable to the lower branch in 
figure \ref{biterrorversusp}) for nearly all instances with noise 
$p_{\rm ch}<0.07$. 

For $p_{\rm ch}=0.07$, we ran our 1RSB algorithm on $20$ instances of a
Sourlas code of $N=999$ bits (note that we need $kN$ to be divisible
by $3$) and compared the results to those of damped time averaged BP (TABP).
Additionally, we compared the results to those of a so-called 
``double loop'' algorithm, which is guaranteed to converge to a local 
minimum of the Bethe free energy \cite{HAK} \footnote{Note that this 
algorithm is 
much more successful for regular lattices, where it provides an implementation
of the cluster variation method. Here it is only used as a comparison since
it will converge to a Bethe free energy minimum even when BP or damped BP 
does not}.
For the sample size of the cavity field distributions we chose $M=1024$, and
we iterated for $50$ updates per field distribution to get rid of transients
before iterating another $50$ updates for the calculation of observables.
These numbers imply a considerable numerical cost, limiting the number of
experiments we could perform in a reasonable amount of time. 
We ran the algorithm for the values $m=0, 0.05, 0.1, 0.15$ and $0.2$ of the
replica symmetry breaking parameter.
TABP was damped with $\epsilon =0.05$, the resulting marginals were averages
over the last $2500$ (sequential) updates of $10000$ in total. For each 
instance we repeated the latter procedure with random initial 
conditions $10$ times.

\subsection{Results}
\begin{figure}
\begin{picture}(400,180)
\put(80,0){\includegraphics[height=6cm, width=8cm]{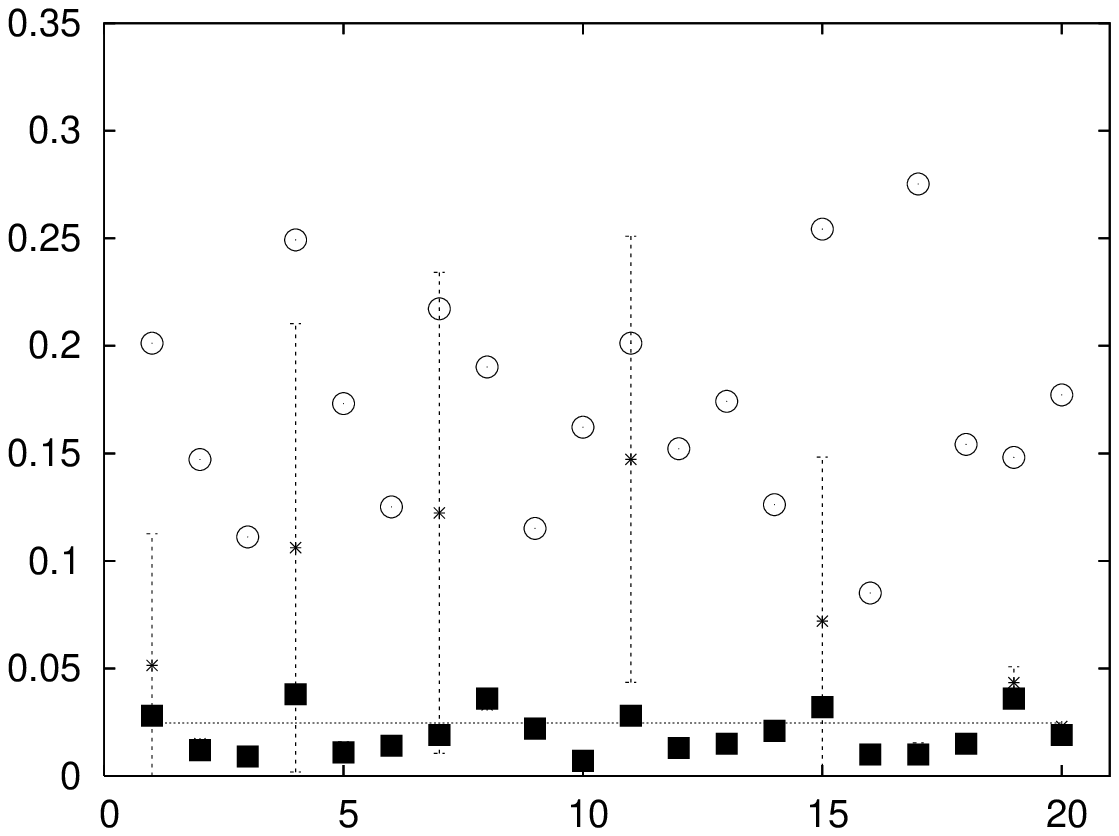}}
\put(65,90){$\rho$}
\end{picture}
\begin{picture}(200,180)
\put(80,10){\includegraphics[height=6cm, width=8cm]{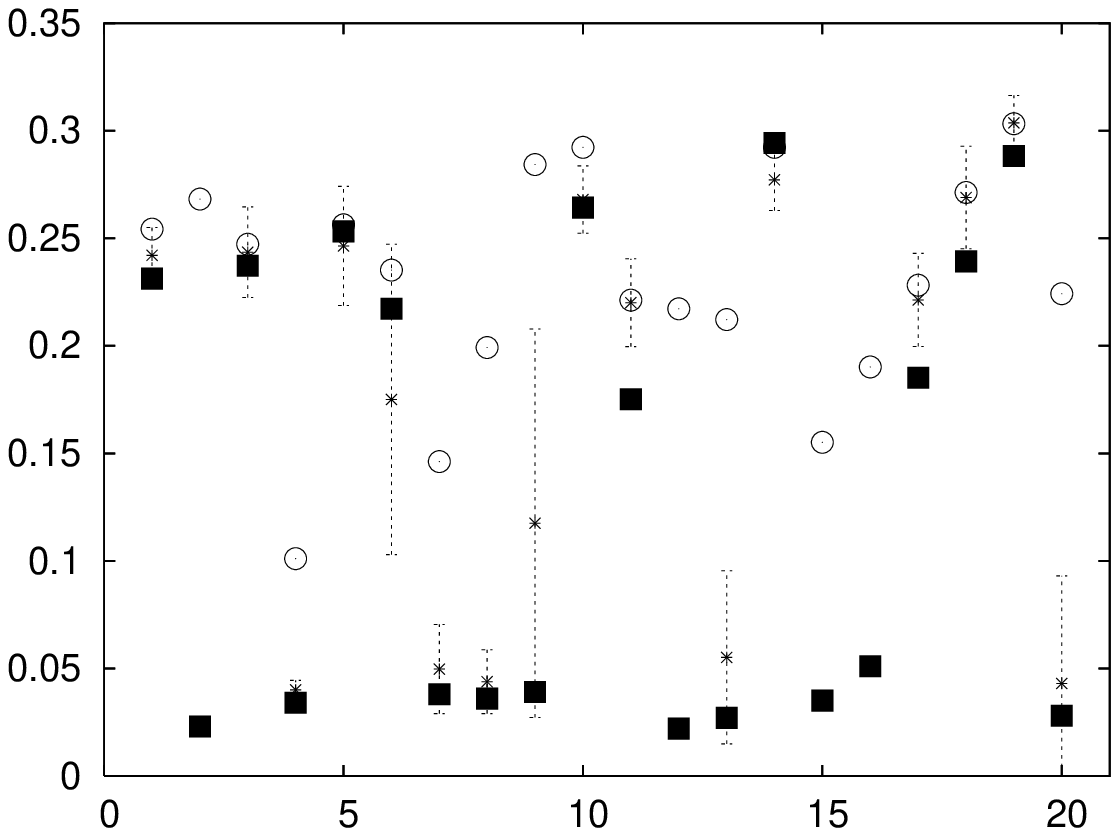}}
\put(65,100){$\rho$}
\put(195,0){instance}
\end{picture}
\caption{$k=4$, $\beta=5$ single instance bit error rate results for 
$20$ systems
of size $N=999$, with channel noise $p_{\rm ch}=0.07$ (top) $p_{\rm ch}=0.08$ 
(bottom). Open circles: double loop algorithm, full squares: our 1RSB 
algorithm, stars with error bars: 10 runs of damped TABP, full square fall
on top when not visible. Top figure 
horizontal line: population dynamics (ferromagnetic initial conditions)
RS result.}
\label{N999pics}
\end{figure}

In figure \ref{N999pics} we report the results of the above experiments for
$p_{\rm ch}=0.07$ and $p_{\rm ch}=0.08$. 
The results of the double loop algorithm, not surprisingly,  
clearly have the largest bit-error
rate, since the algorithm is not designed to find a global
minimum of the Bethe free energy. 
Generically, for small values of the replica symmetry breaking parameter
$m$, the 1RSB algorithm settles in a suboptimal spin glass state, with a 
bit-error rate comparable to the double loop result.
The resulting free energy, which increases with $m$, 
should be maximized with respect to $m$. However, before reaching a maximum,
the free energy drops to the ferromagnetic free energy, 
the corresponding bit-error rate
dropping simultaneously. The optimal value of $m$ is selected by choosing the
smallest $m$ after the free energy drops. 
This behaviour was found for LDPC codes 
macroscopically in \cite{Saadrsb}, but was not yet applied to real instances.

For $p_{\rm ch}=0.07$ 
we observe that our algorithm finds the ferromagnetic state
for each instance at some value of $m$, which still fluctuates between 
different instances due to finite size effects. We conclude that the 
damped TABP algorithm performs surprisingly well on some 
instances. However, whether it
finds a near-optimal state depends on each individual run in general.
The instances $1$, $4$, $7$, $11$, $15$ and $19$ all suffer from initialisation
dependence and have small probability of convergence to the ferromagnetic 
state. The 1RSB algorithm, in contrast, is robust in the sense that it finds
the optimal state for some $m$, regardless of initial conditions. Note that
the 
error-bars of the damped
TABP algorithm are somewhat misleading, since they indicate
the standard deviation, although the distribution of bit-error-rates is not
symmetric.

\begin{figure}
\begin{picture}(150,150)
\put(45,0){\includegraphics[height=5cm, width=6.5cm]{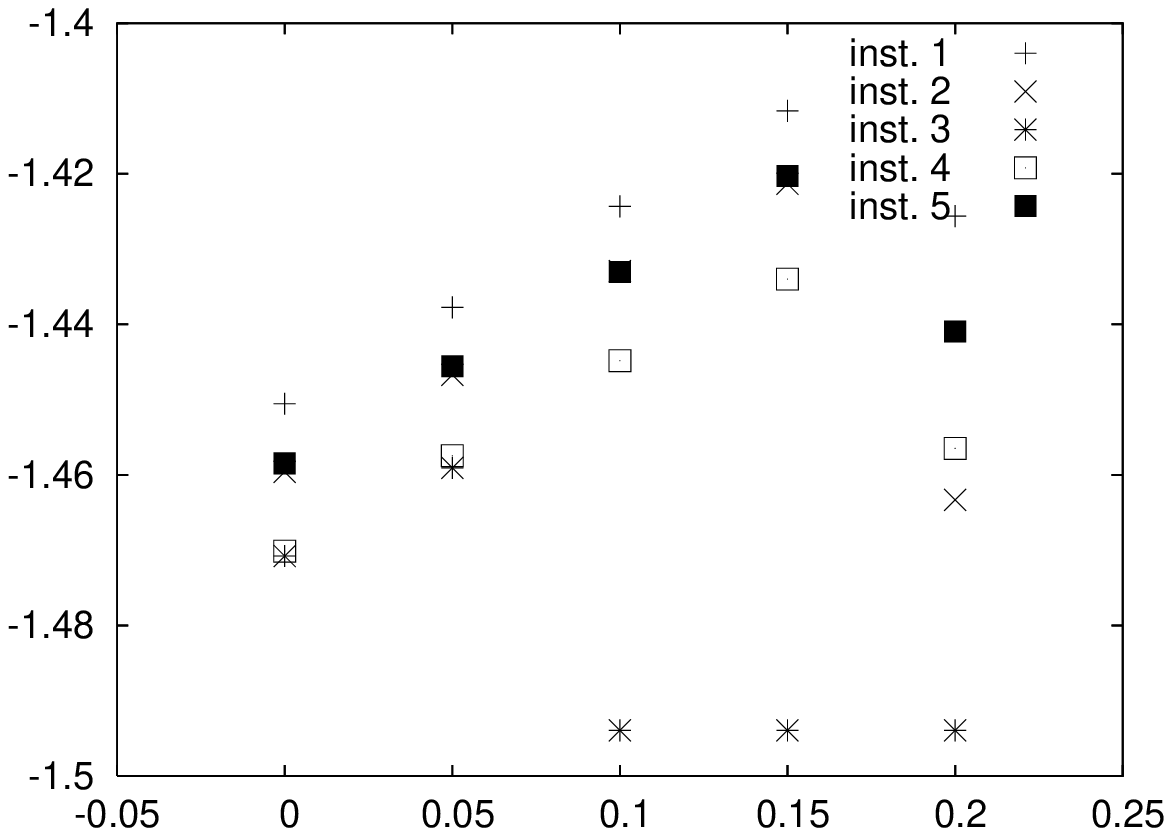}}
\put(30,75){\small{$f(m)$}}
\end{picture}
\begin{picture}(190,150)
\put(-33,0){\includegraphics[height=5cm, width=6.5cm]{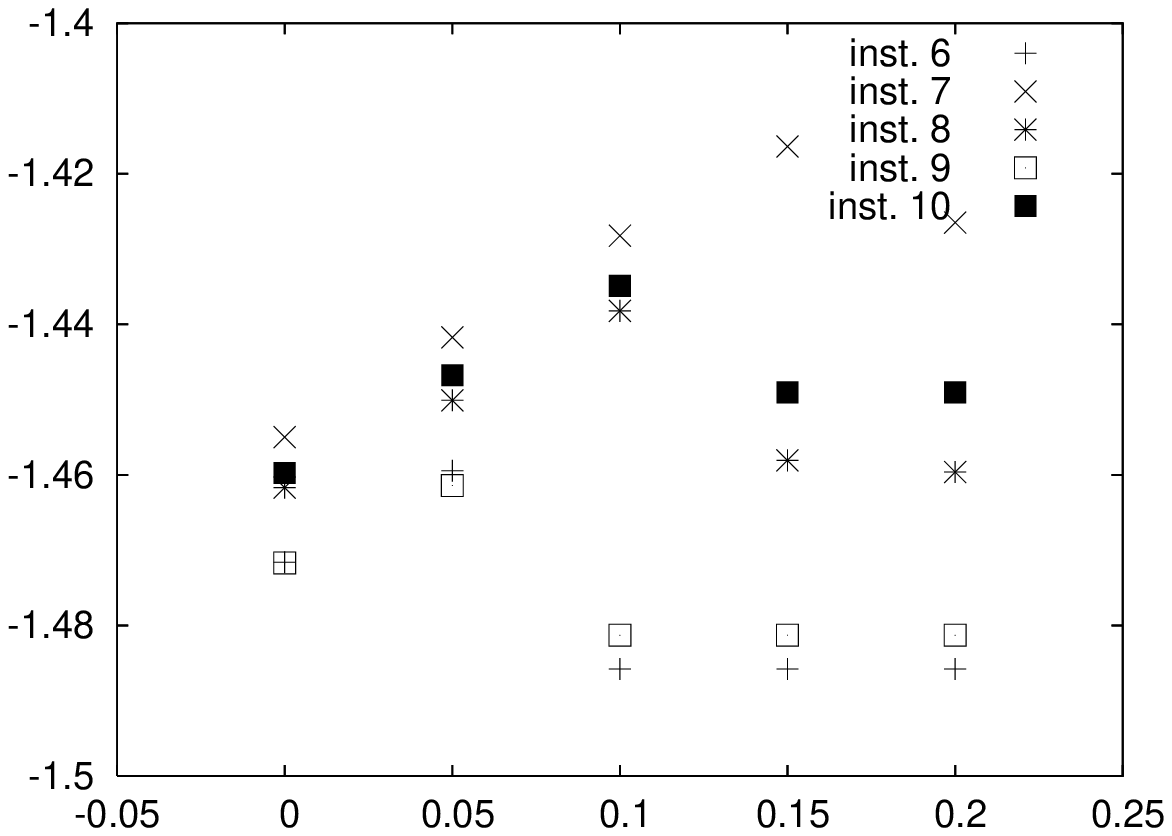}}
\end{picture}
\begin{picture}(190,150)
\put(45,10){\includegraphics[height=5cm, width=6.5cm]{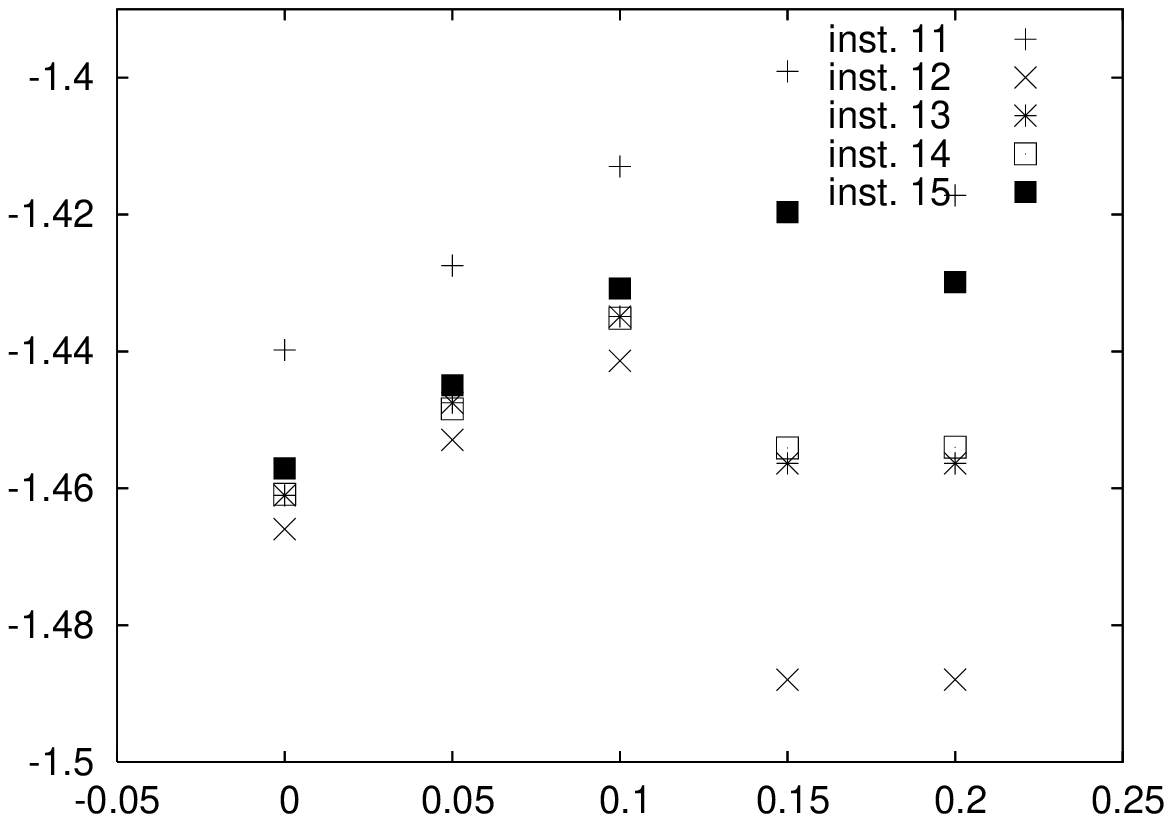}}
\put(30,85){\small{$f(m)$}}
\put(130,0){\small{$m$}}
\end{picture}
\begin{picture}(190,150)
\put(30,10){\includegraphics[height=5cm, width=6.5cm]{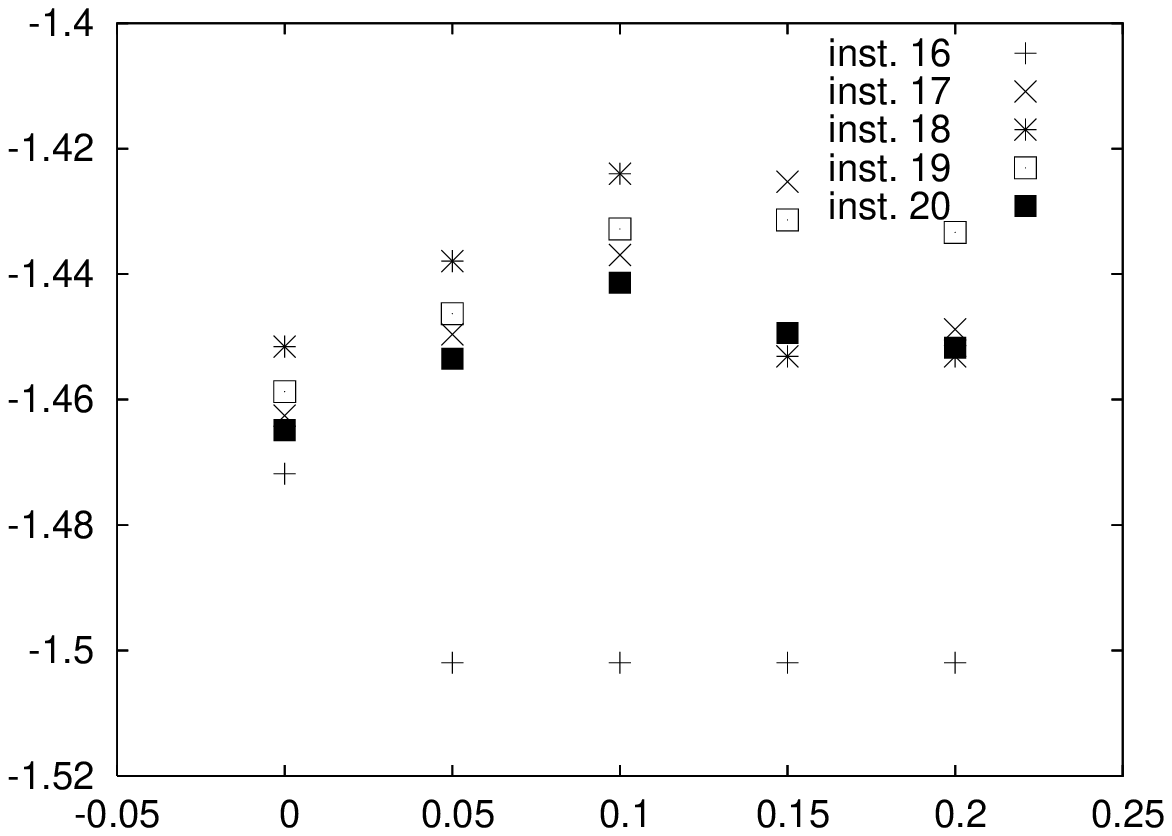}}
\put(130,0){\small{$m$}}
\end{picture}
\caption{Free energy densities as a function of the replica symmetry
breaking parameter $m$ for all instances at
$p_{\rm ch}=0.07$ }
\label{fmfigs}
\end{figure}

In figures \ref{fmfigs} we have plotted all values of the free energy as
a function of the values of $m$ for which we ran the algorithm, illustrating 
the behaviour described above. Note that the maximum of the spin glass free
energy indeed seems to extrapolate to values higher than  
the approximate RS value (around $-1.46$, see figure 
\ref{biterrorversusp}), indicating
that the critical channel noise is likely to be increased in the 1RSB 
framework. 

For $p_{\rm ch}=0.08$, the ferromagnetic state was not always
found by the 1RSB algorithm, indicating either that this is above the 1RSB 
dynamical transition, or we should look in a larger range of $m$.
However, when damped TABP finds a ferromagnetic state, 
in all cases the 1RSB algorithm
finds it robustly. There was one exception to the latter point, in instance
$6$, where TABP found the ferromagnetic state in one of the ten runs. When
we increased the value of $m$ further, however, the algorithm stopped
converging (i.e. the distributions of fields started fluctuating 
heavily). A further increase of the sampling size might be helpful here.

\begin{figure}[t]
\begin{picture}(180,150)
\put(47,10){\includegraphics[height=5cm, width=6cm]{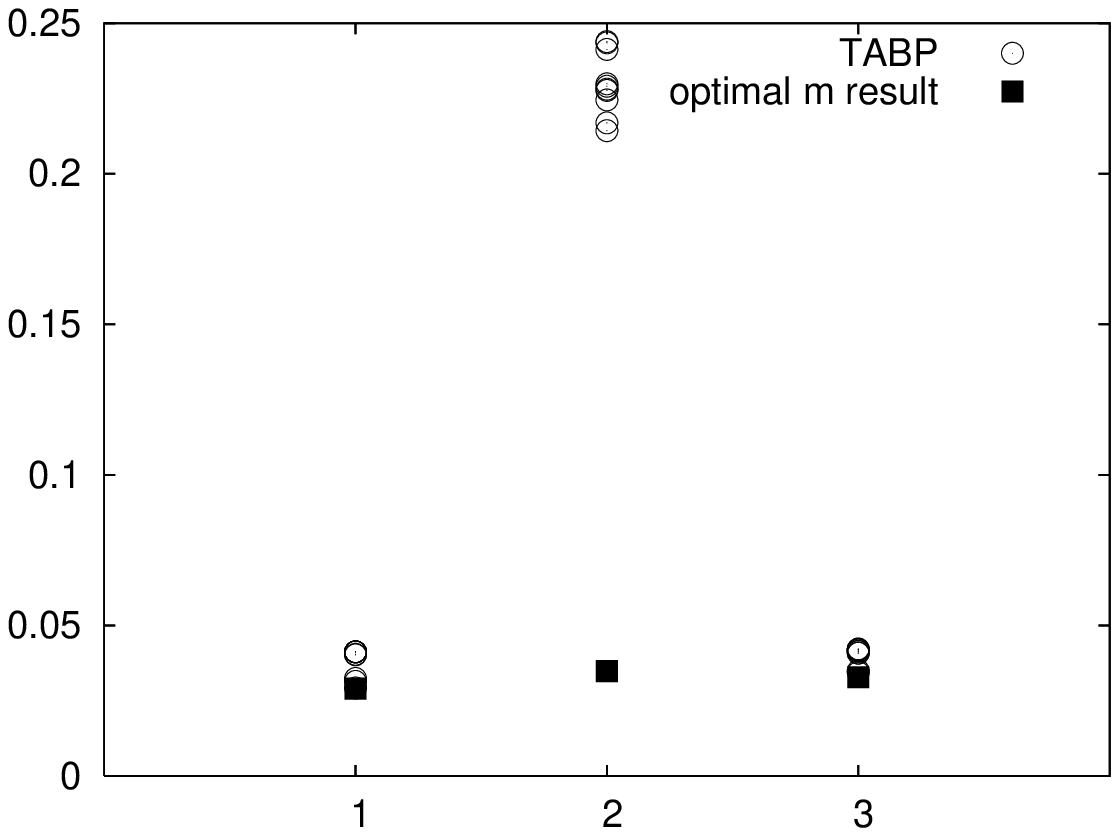}}
\put(30,80){\small{$\rho$}}
\put(110,0){instance}
\end{picture}
\begin{picture}(200,150)
\put(53,10){\includegraphics[height=5cm, width=6cm]{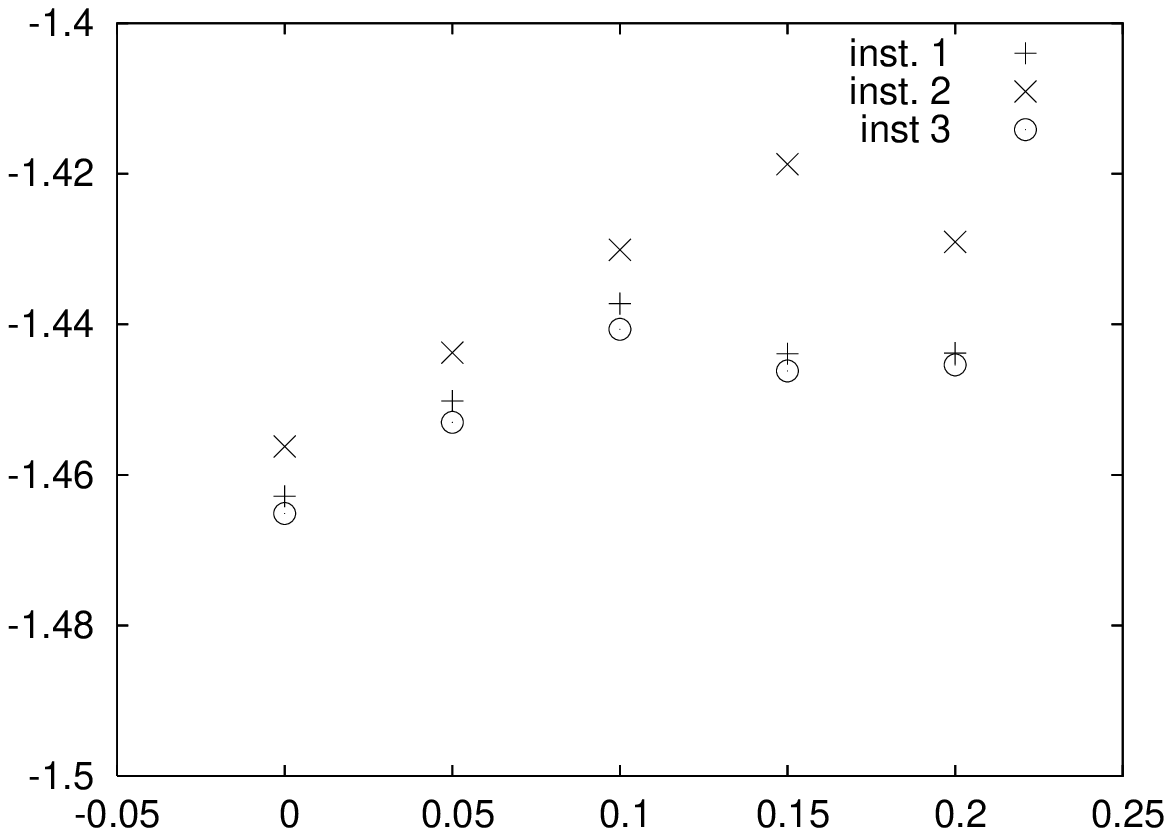}}
\put(35,80){\small{$f(m)$}}
\put(150,0){\small{$m$}}
\end{picture}
\caption{Left: error bit rate comparison between the 1RSB algorithm and $10$ 
runs of TABP at $p_{\rm ch} = 0.075$ and $N=4998$. Right: corresponding
free energy density as a function of $m$.}
\label{figN5000}
\end{figure}

Finally, to reduce finite size effects further, we ran the 1RSB algorithm for 
just $3$ instances for a large code of $N=4998$ at the noise value
$p_{\rm ch}=0.075$. Results are reported in
figures \ref{figN5000}. Still we observe different results for each instance, 
but again the 1RSB algorithm finds the ferromagnetic state in all 3 cases.
This result is spectacular in comparison to the performance of the TABP 
algorithm in
the second instance ,where it did not find the ferromagnetic state in any of 
the $10$ runs.

\section{Discussion}
In this paper we have presented results of applying a finite temperature
generalization of survey propagation on sparse graphs. 
This algorithm, though numerically expensive,
seems particularly of interest for finite temperature decoding
of error correcting codes at temperatures beneath the Nishimori line 
close to the critical channel noise. We have illustrated this by 
regarding a finite connectivity, partly biased 
Sourlas code with $3$-spin interactions. 
As predicted in \cite{Saadrsb}, the dynamical transition of 
belief propagation is pushed towards larger values of the channel noise
for temperatures beneath the Nishimori temperature.
This feature is also reminiscent of zero temperature results obtained for 
survey propagation in the UNSAT phase \cite{ZechBattspy}.

For the model we studied, it turned out that a damped version of 
time averaged belief propagation performed particularly well too. However, the 
finite temperature survey propagation algorithm turned out to be more 
robust, and was able to find the retrieval state in cases where all
other tested algorithms failed.

We feel that this result is an encouragement to further 
investigate the properties, in spite of its numerical cost, of the algorithm, 
which should certainly be tested on the more realistic LDPC codes. 

Furthermore, the dependence of the performance on various parameters, like
the temperature, the bias, and the system size should be investigated 
further. In principle, the simple phase space structure of 
unbiased error correcting codes at the Nishimori
line suggests that in practice it should be easy to infer the value of
the channel noise by varying the temperature in a BP algorithm alone, thereby
circumventing the need for timeconsuming algorithms like the one proposed
in this paper. The present analysis is therefore not intended as the basis 
for a practical
algorithm, but rather as a proof of principle to show (indirectly) that
a finite temperature 1 RSB method can accurately estimate marginal
probabilities on problem instances where BP fails to converge.

Accordingly, the correspondence to results of a macroscopic analysis
of the biased Sourlas code within the 1RSB framework may be further specified,
thereby more carefully specifying the regime in which finite temperature
survey propagation would be an appropriate algorithm.
The latter analysis was omitted in the present study because of 
its numerically extensive nature.

\section*{Acknowledgments}
The authors would like to thank Joris Mooij and Kees Albers for 
stimulating discussions. 
This research was financially 
supported by the Dutch Technology Foundation (STW).

\appendix
\section{Free energy expressions in the 1RSB algorithm}
\label{appfreee}
The expressions for the free energy $F(m)$ and its derivative 
$\partial F(m)/\partial m$ are generalized from \cite{MePa01} in the same
way as the iteration equations. Instead of a collection of $N$ random samples
of fields, constructed by sampling interaction values $J$ and external fields
$\theta$ at each iteration, we now have a distribution of cavity
fields explicitly associated to each link, which is iterated using the 
actual fixed value of
$J$ and $\theta$ corresponding to the instance of the graph under 
consideration. 

Distributions of free energy contributions for $k$ interactions in which
a site is involved
($\Delta F_i^{(1)}$) and for each site linked to $k$ interactions
($\Delta F_i^{(2)}$) are
constructed through the iteration equations
\begin{eqnarray}
\hspace*{-15mm}
-\beta \Delta F_i^{(1)}(\{ J_\mu, h_{\mu_1}, h_{\mu_2}, h_{\mu_3} \}) = 
\sum_{\mu \in V(i)} \log[\cosh(\beta J_\mu)]
\nonumber \\
+ \sum_{\mu \in V(i)}\log[1+\tanh(\beta J_\mu)\tanh(\beta h_{\mu_1})
\tanh(\beta h_{\mu_2}) \tanh(\beta h_{\mu_3})]
\nonumber \\
\hspace*{-15mm}
-\beta \Delta F_i^{(2)}(\{ J_\mu, h_{\mu_1}, h_{\mu_2}\}, \theta_i)
= \log 2 + \sum_{\mu \in V(i)}\log\left\{ \frac{\cosh(\beta J_\mu)}
{\cosh(\beta u(J_{\mu}, h_{\mu_1}, h_{\mu_2}))} \right\}
\nonumber \\
+ \log \cosh\left\{ \beta \left[ \sum_{\mu \in V(i)} u(J_{\mu}, 
h_{\mu_1}, h_{\mu_2}) + \theta_i \right] \right\}
\end{eqnarray}
The contribution of each site to the total free energy is (taking into
account overcounting of link contributions)
\begin{equation}
\Delta F_i = - \frac{2}{3} \Delta F_i^{(1)} + \Delta F_i^{(2)}
\end{equation}
in a replica symmetric setting. In the 1RSB framework, these quantities
should be averaged over all states, weighted with their Boltzmann factors 
as in 
\cite{MePa01} and this result is then averaged over the distribution of
free energies. 
The resulting expressions that are evaluated numerically are
\begin{equation}
f_i = -\frac{2}{3} F_i^{(1)} + F_i^{(2)} 
\end{equation}
with 
\begin{eqnarray}
F_i^{(1)} = -\frac{1}{m \beta} \log \frac{1}{M} \sum_\alpha 
e^{-m\beta \Delta F_i^{(1)\alpha}} \nonumber \\
F_i^{(2)} = -\frac{1}{m \beta} \log \frac{1}{M} \sum_\alpha 
e^{-m\beta \Delta F_i^{(2)\alpha}}
\end{eqnarray}

\section{Cavity analysis of Sourlas code}
\label{appsourlas}
In the following it will be useful to introduce 
sublattices: $I_\alpha = \{ i | \theta_i = \theta_\alpha\}$
to take into account the fact that the sites are not 
equivalent due to their biases. The order parameter will thus be
the collection of three sublattice distributions of cavity fields, given the
message:
$W_+(h|\xi)$, $W_-(h|\xi)$ and $W_0(h|\xi)$. The order parameter equations 
are given by
\begin{eqnarray}
\hspace*{-15mm}
W_\alpha(h|\xi) =
\int \prod_{i=1}^{k-1}[d u_i Q_\alpha(u_i|\xi)]
\delta[h-\sum_{i=1}^{k-1} u_i - \theta_\alpha] \nonumber\\
\hspace*{-15mm}
Q_\alpha(u|\xi) = 
\sum_{\alpha', \alpha''} c(\alpha')c(\alpha'')
\sum_J \sum_{\xi', \xi''}
p(J|\xi, \xi',\xi'')p(\xi',\xi''|\alpha',\alpha'') \nonumber \\
\hspace*{-10mm}
\times \int dh dg W_{\alpha'} (h|\xi') W_{\alpha''}(g|\xi'')
\delta[u - \frac{1}{\beta}\tanh^{-1}[\tanh(\beta h)\tanh(\beta g)
\tanh(\beta J)]] \nonumber \\
\qquad
\end{eqnarray}
or
\begin{eqnarray}
\hspace*{-15mm}
Q_\alpha(u|\xi) = 
\sum_{\alpha', \alpha''} c(\alpha')c(\alpha'')
\sum_J \sum_{\xi', \xi''}
\frac{e^{\beta_p[J\xi \xi'\xi'' + \xi' \theta_{\alpha'} + \xi'' \theta_{\alpha''}]}}
{8 \cosh(\beta_p)\cosh(\beta_p \theta_{\alpha'})\cosh(\beta_p \theta_{\alpha''})}
\nonumber \\
\hspace*{-10mm}
\times \int dh dg W_{\alpha'} (h|\xi') W_{\alpha''}(g|\xi'')
\delta[u - \frac{1}{\beta}\tanh^{-1}[\tanh(\beta h)\tanh(\beta g)
\tanh(\beta J)]] \nonumber \\
\qquad
\end{eqnarray}
These equations can be simplified by considering symmetries.
First of all, we recognize that $Q_\alpha(u|\xi)$ does not actually depend on 
$\alpha$, and obeys $Q(u|-\xi)=Q(-u|\xi)$. From this symmetry it follows that
$W_\alpha(h|-\xi) = W_{-\alpha}(-h|\xi)$. As a result we may restrict
to considering either $W_\alpha(h|+)$ or $W_{\alpha}(h|-)$. We will assume that
$c(+) = c(-)$ from now on. This gives us
\begin{eqnarray}
\hspace*{-15mm}
Q(u|+)  = 
\sum_J  \frac{e^{\beta_p J}}{2 \cosh(\beta_p)} 
\nonumber \\
\hspace*{-10mm}
\int dh dg \left[ c(0)
W_0 (h|+)
  + c(+) \frac{e^{\beta_p\theta}}{\cosh(\beta_p \theta)}  
W_+(h|+) + c(-)  \frac{e^{-\beta_p \theta}}{\cosh(\beta_p \theta)}  
W_-(h|+) \right] \nonumber \\ 
\times\left[ c(0)
W_0 (g|+) + c(+) \frac{e^{\beta_p\theta}}{\cosh(\beta_p \theta)}  
W_+(g|+) + c(-)  \frac{e^{-\beta_p \theta}}{\cosh(\beta_p \theta)}  
W_-(g|+) \right]
\nonumber \\
\times ~
\delta\left[u - \frac{1}{\beta}\tanh^{-1}[\tanh(\beta h)\tanh(\beta g)
\tanh(\beta J)]\right] 
\end{eqnarray}
We are interested in the overlap with
the message at the site in question. Therefore we regard the joint 
distribution of field and message:
\begin{equation}
\hspace*{-20mm}
W_\alpha(H,\xi) =  p(\xi|\theta_\alpha)W_\alpha(H|\xi) = p(\xi|\theta_\alpha) 
\int \prod_{i=1}^k[d u_i Q_\alpha(u_i|\xi)]
\delta[H-\sum_{i=1}^k u_i - \theta_\alpha] 
\end{equation}
If the sign of $\xi_i \tanh(\beta H_i)$ is positive, the message at site
$i$ will be decoded correctly. Thus the overall bit error rate is $(1-\mu)/2$, 
where
\begin{equation}
\mu = \frac{1}{2}\sum_\xi \sum_{\alpha} c(\alpha) \int d H 
W_\alpha(H,\xi) {\rm sgn}(\xi H)
\end{equation}
Since $p(\xi|-\theta_\alpha) = p(-\xi|\theta_\alpha)$ and
$W_{\alpha}(H|-\xi) = W_{-\alpha}(-H|\xi)$, we may deduce 
$W_{\alpha}(H,-\xi) = W_{-\alpha}(-H,\xi)$ and consequently
\begin{eqnarray}
\hspace*{-20mm}
\mu = \int dH \Biggl[c(0) W_0(H|+) \nonumber \\
 +  ~ c(+) \frac{e^{\beta_p \theta}}
{\cosh(\beta_p \theta)} W_+(H|+) +  c(-) \frac{e^{-\beta_p \theta}}
{\cosh(\beta_p \theta)} W_-(H|+) \Biggr]{\rm sgn}(H)
\qquad
\end{eqnarray}
All equations clearly close in terms of $Q(u)$, $W(h)$ and $W(H)$, where
\begin{eqnarray}
Q(u) = Q(u|+) \nonumber \\
W(h) = \sum_{\alpha} q(\theta_\alpha) W_\alpha(h|+) \nonumber \\
W(H) = \sum_{\alpha} q(\theta_\alpha) W_\alpha(H|+)
\end{eqnarray}
with the effective measure
\begin{equation}
q(\theta_\alpha) = c(\alpha) \frac{e^{\beta_p \theta_\alpha}}{\cosh(\beta_p 
\theta_\alpha)}
\end{equation}
leading to equations (\ref{caviter}).
\end{document}